\documentclass[aps,twocolumn,superscriptaddress]{revtex4}
\usepackage{epsfig}
\usepackage{times}
\usepackage{color}

\begin{document}

\title{Wisdom of groups promotes cooperation in evolutionary social dilemmas}

\author{Attila Szolnoki}
\email{szolnoki.attila@ttk.mta.hu}
\affiliation{Institute of Technical Physics and Materials Science, Research Centre for Natural Sciences, Hungarian Academy of Sciences, P.O. Box 49, H-1525 Budapest, Hungary}

\author{Zhen Wang}
\affiliation{School of Physics, Nankai University, Tianjin 300071, China}
\affiliation{Department of Physics, Hong Kong Baptist University, Kowloon Tong, Hong Kong}

\author{Matja{\v z} Perc}
\email{matjaz.perc@uni-mb.si}
\affiliation{Faculty of Natural Sciences and Mathematics, University of Maribor, Koro{\v s}ka cesta 160, SI-2000 Maribor, Slovenia}

\begin{abstract}
Whether or not to change strategy depends not only on the personal success of each individual, but also on the success of others. Using this as motivation, we study the evolution of cooperation in games that describe social dilemmas, where the propensity to adopt a different strategy depends both on individual fitness as well as on the strategies of neighbors. Regardless of whether the evolutionary process is governed by pairwise or group interactions, we show that plugging into the ``wisdom of groups'' strongly promotes cooperative behavior. The more the wider knowledge is taken into account the more the evolution of defectors is impaired. We explain this by revealing a dynamically decelerated invasion process, by means of which interfaces separating different domains remain smooth and defectors therefore become unable to efficiently invade cooperators. This in turn invigorates spatial reciprocity and establishes decentralized decision making as very beneficial for resolving social dilemmas.
\end{abstract}

\maketitle

The evolutionary success of the human species is to a large extent due to our exceptional other-regarding abilities \cite{nowak_11}. Cooperation in the genus \textit{Homo} is believed to have evolved to help us cope with rearing offspring that survived \cite{hrdy_11}, as well as to help us mitigate between-group conflicts \cite{bowles_11}. Today it is instrumental for harvesting the benefits of collective efforts on an unprecedented scale, which in turn fuels the progress and innovation we are able to witness on a daily basis. However, it may well be that these behavioral predispositions also make us very susceptible to what others are doing, as well as saying and even thinking about us \cite{asch_sa55}. Everyone fancies being unique, yet the reality is such that much of our individuality is lost to conformance in both appearance and behavior. On the other hand, tuning in to the masses also has notable advantages, most prominently creating the opportunity to exploit the so-called ``wisdom of the crowd'' effect \cite{surowiecki_04}. Indeed, the first evidence suggesting that the average of many estimates may be closer to the truth as individual, albeit expert opinions is more than a century old \cite{galton_n1907}. In this paper, we investigate the merit of this fascinating phenomenon for the successful evolution of cooperation in games that describe social dilemma situations \cite{hofbauer_98, nowak_06, sigmund_10}.

Evolutionary games can be classified to those that are governed by pairwise interactions and to those that are governed by group interactions. The prisoner's dilemma game \cite{axelrod_84} is a classical example of an evolutionary game that is governed by pairwise interactions. The public goods game, on the other hand, is a typical example of an evolutionary game that is governed by group interactions. As highlighted by recent research \cite{szolnoki_pre09c, wu_t_pre09, wu_t_epl09, perc_njp11, gomez-gardenes_c11, gomez-gardenes_epl11, santos_pnas11, arenas_jtb11, van-segbroeck_prl12, vukov_njp12, cardillo_pre12, li_j_pa12, wu_t_pre12, gracia-lazaro_srep12}, the distinction may be crucial as group interactions provide an inherently different environment, one where the trails of those that defect are blurred and where therefore proper reciprocity is a challenge. Regardless of the distinction, when individual prosperity calls for actions that harm the wellbeing on the collective level, the population is faced with a social dilemma that may results in a ``tragedy of the commons'' \cite{hardin_g_s68}. While several mechanisms that prevent defection from taking over have already been discovered \cite{nowak_s06}, the identification of conditions for the survival and spreading of cooperation among selfish individuals still remains a grand challenge, both by games governed by pairwise as well as by games governed by group interactions.

Recent reviews are a testament to the vibrancy of this field of research \cite{szabo_pr07, schuster_jbp08, roca_plr09, perc_bs10}, linking together knowledge from biology, sociology, economics as well as mathematics and physics, to provide a better understanding of why we so frequently choose cooperation over defection. Particularly relevant for the present work are preceding papers considering different learning abilities of players \cite{szolnoki_njp08, chen_xj_ijmpc08, zhang_hf_pa10b, szolnoki_pre10b, tanimoto_pre12, dai_ql_pre12}, where however the latter were assumed by default and altered by explicit rules rather than being something that emerges and varies spontaneously in dependence on the behavior of the neighbors. Most importantly, here we assume that the propensity of each individual to adopt a new strategy depends on the behavior within the group of immediate neighbors. If the strategy of the player is the same as that of the majority of other players its learning activity will be low. Conversely, if the strategy is different than that of the majority, the learning activity of the player will be high. True to a physicist approach, a single parameter $\alpha$ interpolates between completely ignoring the ``wisdom of the group'' and considering it ardently. Note that the assumption is that the strategy of each individual is representative for its knowledge, and that thus the collective information on the strategies within the group is equally representative for the group wisdom. Thereby we thus wed the ``wisdom of the crowd'' effect \cite{surowiecki_04} with evolutionary games, albeit referring to groups rather than crowds to adjust the terminology accordingly.

As we will show in what follows, the stronger the feedback between the group and the individual player, i.e., the larger the value of $\alpha$ (see Fig.~\ref{alpha} and the Methods section), the more successful the evolution of cooperation. This holds true for both the prisoner's dilemma and the public goods game, thus pointing towards a universal mechanism that bridges the distinction between games governed by pairwise and games governed by group interactions. Responsible for this is a dynamically modified invasion process that is completely strategy independent, yet by means of which cooperators, unlike the defectors, are able to draw a significant evolutionary advantage. On the one hand, it helps cooperators to build sizable compact clusters with smooth interfaces, and on the other, it prevents defectors to invade rapidly. The fact that retarding fast invasion has unequal consequence for the competing strategies has been emphasized before in the context of aging in evolutionary games \cite{szolnoki_pre09} and the evolution of public cooperation on interdependent networks \cite{wang_z_epl12}, which altogether points to a discovery of a general and widely applicable phenomenon, during which the ``wisdom of groups'' is exploited for the effective resolution of social dilemmas. More detailed results will be presented next, while for details concerning the employed evolutionary games we refer to the Methods section.

\section*{Results}

\begin{figure}
\centerline{\epsfig{file=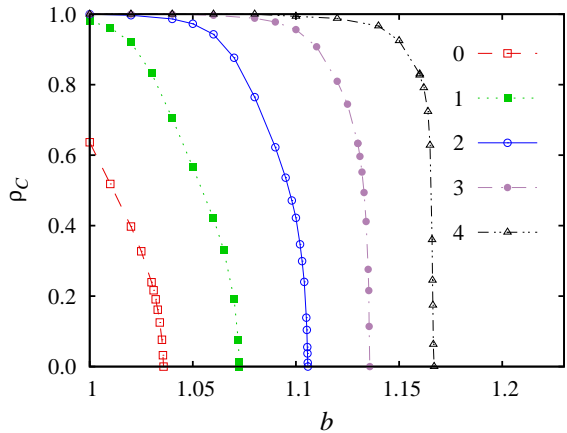,width=8.4cm}}
\caption{\label{pdsq} The ``wisdom of groups'' promotes the evolution of cooperation in the prisoner's dilemma game on the square lattice. Depicted is the fraction of cooperators $\rho_C$ in dependence on the temptation to defect $b$, as obtained for different values of $\alpha$ (see figure legend). It can be observed that the larger the value of $\alpha$ the higher the value of $b$ where cooperators are able to dominate and survive the evolutionary competition with defectors.}
\end{figure}

We begin by presenting results for the prisoner's dilemma game on the square lattice. Figure~\ref{pdsq} shows how the stationary fraction of cooperators $\rho_C$ varies in dependence on the temptation to defect $b$ for different values of $\alpha$. The curve for $\alpha=0$ is a well-known result of evolutionary spatial games, according to which in the absence of the ``wisdom of the group'' effect cooperators on the square lattice die out at the critical $b(K=0.1)=1.0358$, and the extinction  belongs to the directed percolation universality class \cite{szabo_pre05}. As the knowledge of the group is taken into account, however, the critical temptation to defect shifts towards larger values. In fact, the higher the value of $\alpha$, the higher the critical value of $b$ at which cooperators die out. Also worth highlighting is that for $\alpha=0$ cooperators are never able to completely dominate the population. Conversely, for $\alpha>1$ there always exist a critical temptation to defect below which defectors die out. Again, the higher the value of $\alpha$, the higher the value of $b$ at which defectors are still unable to survive. The conclusion that imposes itself is thus that the ``wisdom of groups'' promotes the evolution of cooperation and thus aids the resolution of social dilemmas. Not only are the cooperators able to survive at higher values of $b$, but also their range of complete dominance increases with increasing values of $\alpha$.

\begin{figure}
\centerline{\epsfig{file=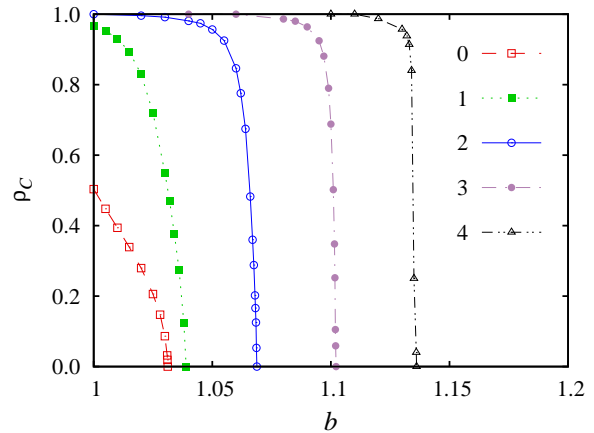,width=8.4cm}}
\caption{\label{pdhon} The ``wisdom of groups'' promotes the evolution of cooperation irrespective of the properties of the interaction network. In this figure the prisoner's dilemma game was staged on the honeycomb lattice. Depicted is the fraction of cooperators $\rho_C$ in dependence on the temptation to defect $b$, as obtained for different values of $\alpha$ (see figure legend). As by the results obtained on the square lattice depicted in Fig.~\ref{pdsq}, it can be observed that the larger the value of $\alpha$ the higher the value of $b$ where cooperators outperform defectors.}
\end{figure}

Further supporting the generality of our observations are results obtained on the honeycomb lattice, which are presented in Fig.~\ref{pdhon}. The honeycomb lattice has the lowest degree possible for spatial graphs ($z=3$), which should minimize the effects of spatial reciprocity. Despite of this, taking into account the ``wisdom of groups'' has virtually an identical impact on the evolution of cooperation as it does on the square lattice. It is also well known that cooperation is markedly affected by the type of the interaction network \cite{szabo_pr07}, where in particular the scale-free degree distribution has been identified as a potent promoter of cooperative behavior \cite{santos_prl05, santos_pnas06, poncela_njp07, gomez-gardenes_prl07, rong_pre07, szolnoki_epl08, assenza_pre08, perc_njp09, pena_pre09, galan_pone11, yang_hx_epl12, allen_jtb12, buesser_pre12, pinheiro_njp12}. Furthermore, more subtle properties of interaction networks, such as the clustering coefficient and the average degree of players have also been emphasized as being crucial \cite{vukov_pre06}. We therefore simulated the evolutionary process also on the triangular lattice, which has a high clustering coefficient, and we observed a qualitatively similar impact of the ``wisdom of groups'' effect. Thereby we thus establish that the reported promotion of cooperation is to a large degree independent of the particularities of the interaction network.

\begin{figure}
\centerline{\epsfig{file=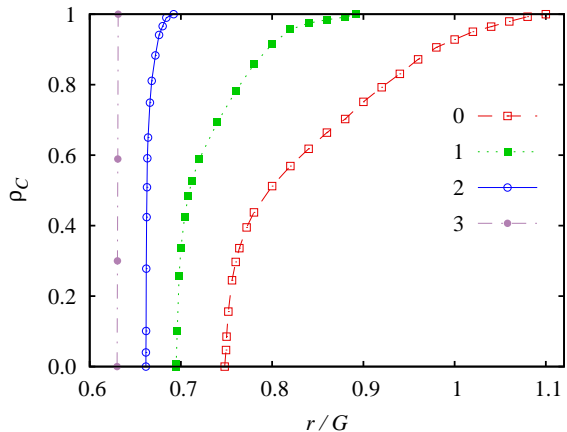,width=8.4cm}}
\caption{\label{pggsq} The ``wisdom of groups'' promotes the evolution of cooperation irrespective of the type of the game. In this figure the public goods game was staged on the square lattice. Depicted is the fraction of cooperators $\rho_C$ in dependence on the multiplication factor $r$, as obtained for different values of $\alpha$ (see figure legend). As by the results obtained for the prisoner's dilemma game in Figs.~\ref{pdsq} and \ref{pdhon}, it can be observed that the larger the value of $\alpha$ the lower the multiplication factor $r$ that is needed for cooperators to survive and dominate the population.}
\end{figure}

Besides different interaction networks, the distinction between games governed by pairwise interactions and games governed by group interactions may also play a crucial role, as emphasized in the Introduction. While the prisoner's dilemma is the paradigmatic example of a pairwise driven game, the same is true for the public goods game as the game that is governed by group interactions. With this in mind, we present in Fig.~\ref{pggsq} the stationary fraction of cooperators in dependence on the multiplication factor $r$ for different values of $\alpha$. Note that unlike for the value of $b$ in the prisoner's dilemma game, here the lower the value of $r$ the more challenging the evolution of cooperation. For $\alpha=0$ the outcome is again a well-known result, according to which cooperators die out at the critical $r/G=0.748$ and start dominating the population completely at $r/G=1.1$ \cite{szolnoki_pre09c}. However, for larger values of $\alpha$ the ``wisdom of groups'' is introduced to the players, and accordingly the critical values of $r$ shift. It can be observed that the higher the value of $\alpha$, the lower the critical multiplication factors at which cooperators are able to survive and dominate the evolutionary competition with defectors. It is also worth pointing out that the interval of $r$ values supporting the coexistence of both strategies shrinks continuously as $\alpha$ increases. Altogether, these results convincingly support the fact that the ``wisdom of groups'' effect supports the evolution of cooperation not only in games governed by pairwise interactions, but also in games that are governed by group interactions.

\begin{figure*}
\centerline{\epsfig{file=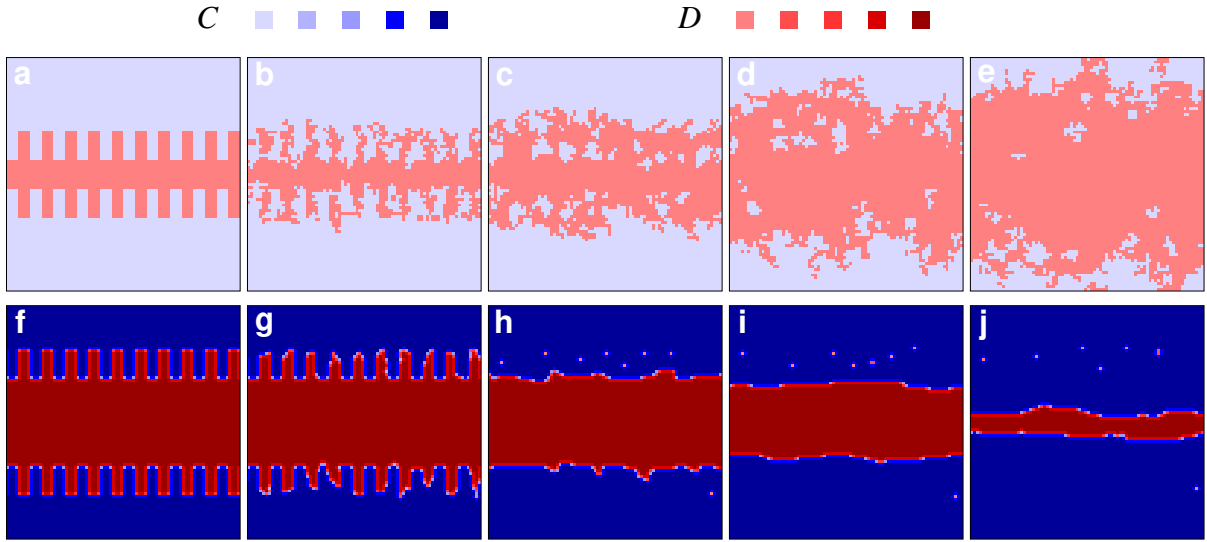,width=16cm}}
\caption{\label{prepared} Interfaces that separate domains of cooperators and defectors remain smooth if the ``wisdom of groups'' is taken into account. Depicted is the evolution of a prepared initially rough interface, as obtained for $\alpha=0$ [top panels from (a) to (e)], i.e., ignoring the ``wisdom of groups'', and for $\alpha=4$ [bottom panels from (f) to (j)]. It can be observed that by taking into account the wider knowledge of the group (bottom row), the roughening of the interfaces is prevented. Cooperators therefore rise to dominance (the final pure $C$ phase is not shown). In the upper row the defectors are able to invade the cooperative domains aggressively, which in turn further roughens the interfaces and eventually leads to a pure $D$ phase (not shown). Note that to distinguish different learning activities $w$ of players, we have used different shades of blue (for cooperators) and red (for defectors), as indicated in the figure legend. Lighter colors correspond to higher learning activity while darker colors denote players with lower learning activity. For $\alpha=0$ all players constantly have $w=1$, and are accordingly depicted by the brightest shades of blue and red. The snapshots were taken at MCS=$0$, $10$, $30$, $70$ and $100$ for the top row, and at MCS=$0$, $200$, $1400$, $6000$ and $19000$ for the bottom row. In both cases the temptation to defect was set equal to $b=1.07$ and the system size was $L=80$ (small solely to ensure a proper resolution of the relevant spatial patterns).}
\end{figure*}

Given the general applicability of the reported cooperation-promoting mechanism, it is of significant interest to elucidate why then taking into account the wider knowledge of groups upon deciding whether or not to change a strategy is so beneficial for the effective resolution of social dilemmas. Henceforth we focus on the prisoner's dilemma game on the square lattice, although we note that qualitatively identical results can also be obtained on other lattices or with the public goods game.

First, we present in Fig.~\ref{prepared} characteristic spatial patterns that emerge if the evolutionary process is initiated from a prepared initial state. The goal is to start with rough interfaces, as the evolution along them has often been identified as crucial for deciding the winner between cooperators and defectors in spatial games. For $\alpha=0$ (top row), in the absence of the ``wisdom of groups'' where the learning activity of all players is the same (see Fig.~\ref{alpha}), defectors can easily invade cooperators, and their invasions make the interface increasingly irregular [panel (b)]. This in turn further weakens spatial reciprocity and finally results in a pure $D$ phase (not shown). For $\alpha=4$ (bottom row), when the ``wisdom of groups'' is turned on and accordingly the learning activity of players is strongly driven by the strategies of their neighbors, the evolution is significantly different. Here instead of additional roughening due to invading defectors, the interface is actually smoothed as the ``peaks'' of defectors are rapidly suppressed [panels (g) and (h)]. The result of this process are smooth straight interfaces that bolster the effectiveness of spatial reciprocity, and hence enable the survival of cooperators even at large temptations to defect, as demonstrated in Fig.~\ref{pdsq}. We note that in the bottom row of Fig.~\ref{prepared} cooperators eventually come to dominate the population completely (not shown). Interestingly, the promotion of cooperation due to the smoothing of interfaces has recently been reported also in the context of the spatial public goods game with adaptive punishment \cite{perc_njp12}. Here, however, the reason lies not in punishing defectors, but rather in a dynamically modified invasion process, as we will further explain in what follows.

\begin{figure}
\centerline{\epsfig{file=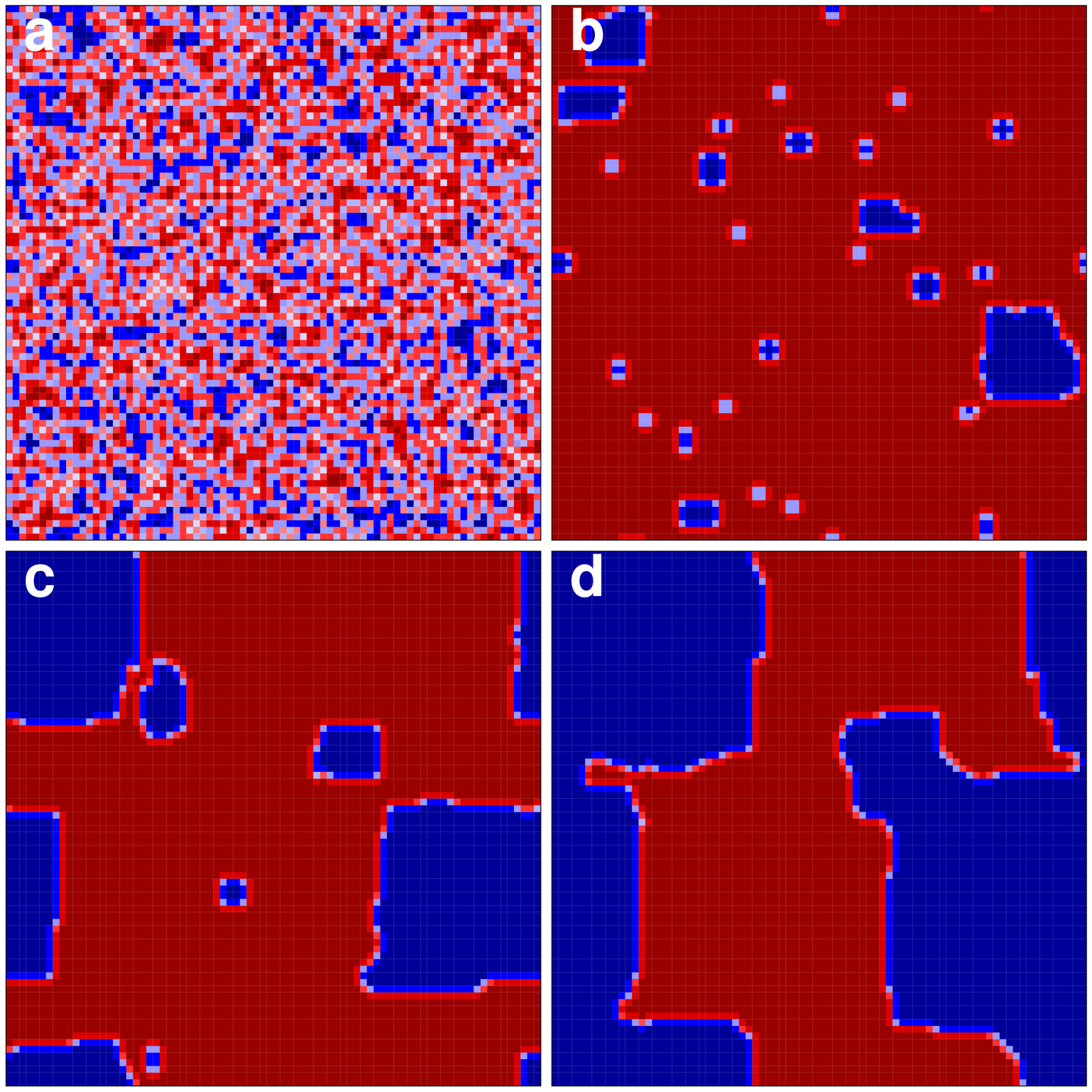,width=8.5cm}}
\caption{\label{random} Evolution of cooperation from a random initial state under the influence of the ``wisdom of groups''. Depicted are characteristic spatial patterns, as obtained for $\alpha=4$ and $b=1.08$. The color code is the same as used in Fig.~\ref{prepared}. Defectors with the highest learning activity, depicted light red, can be observed at the interfaces that separate ordered domains. Cooperators are able to overtake these defectors, which results in smoother interfaces and ultimately in a pure $C$ phase (not shown). Also note that only the cooperative domains with ``straight'' borders and of sufficiently large size are able to prevail against the invading defectors. Smaller circularly shaped cooperative domains are unable to grow and surrender to the evolutionary pressure rather fast. The snapshots were taken at MCS=$0$ (a), $10000$ (b), $30000$ (c) and $41000$ (d). The system size was $L=80$.}
\end{figure}

Equipped with the insight from the evolution of spatial patterns from a prepared initial state as presented in Fig.~\ref{prepared}, it is next of interest to investigate the evolution of spatial patterns under the effect of the ``wisdom of groups'' from a random initial state. Figure~\ref{random} features a series of snapshots that are characteristic for an evolutionary process that is subject to spatial reciprocity. Cooperators are first rather quickly brought to the brink of extinction [panel (b)], only to recuperate rather spectacularly after properly regrouping and forming sufficiently large compact clusters [panels (c) and (d)]. The final state is a pure $C$ phase, which is not shown. It is important to note that the ``wisdom of groups'' invigorates spatial reciprocity, yet it also requires a more delicate formation of cooperative clusters to begin with. We emphasize that the majority of small circular $C$ domains in panel (b) vanishes. Only those with ``straight'' borders and of sufficient size form the nucleus from which the reinvasion of cooperators can begin, as depicted in panels (c) and (d). The deciding battle between the two competing strategies thus takes place along the interfaces where there are ``steps'' in the straight fronts. Here, players of both strategies are ``vulnerable'' and susceptible to strategy change because they have relatively high learning activities (these players are marked with brighter red and blue colors), while away from such steps the evolutionary process is significantly decelerated. Although this effect affects as much defectors as it does cooperators, the later are able to draw an evolutionary advantage. The promotion of cooperation thus relies on the spontaneous emergence of a dynamically decelerated invasion process in the vicinity of straight interfaces. This is reminiscent of the effect of aging \cite{szolnoki_pre09} as well as interdependent networks \cite{wang_z_epl12}, although here it emerges spontaneously due to the ``wisdom of groups'' effect.

\begin{figure}
\centerline{\epsfig{file=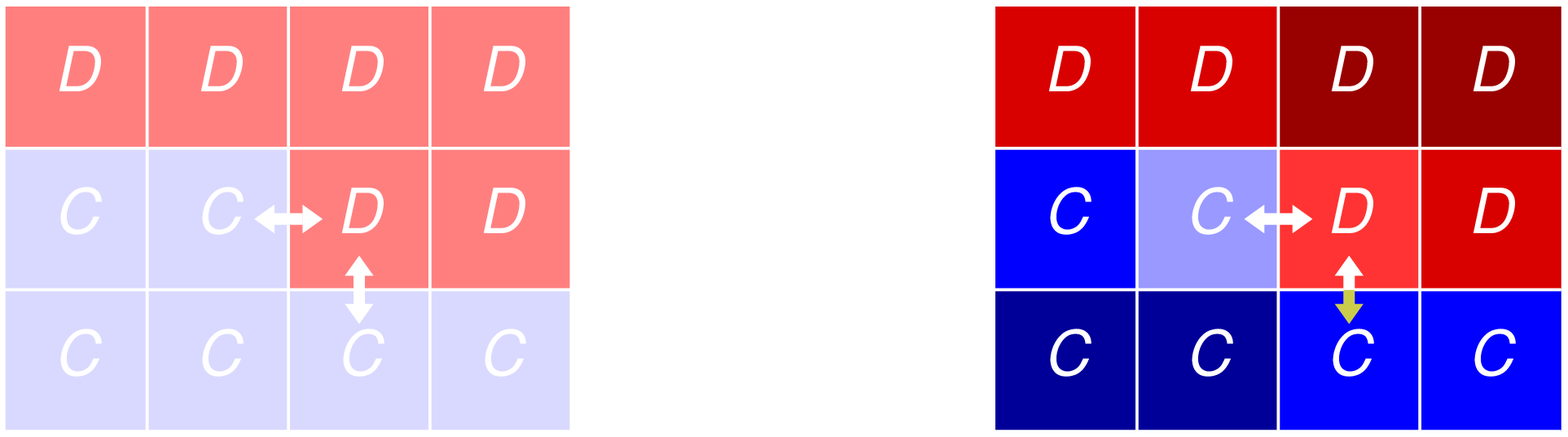,width=8.4cm}}
\centerline{\epsfig{file=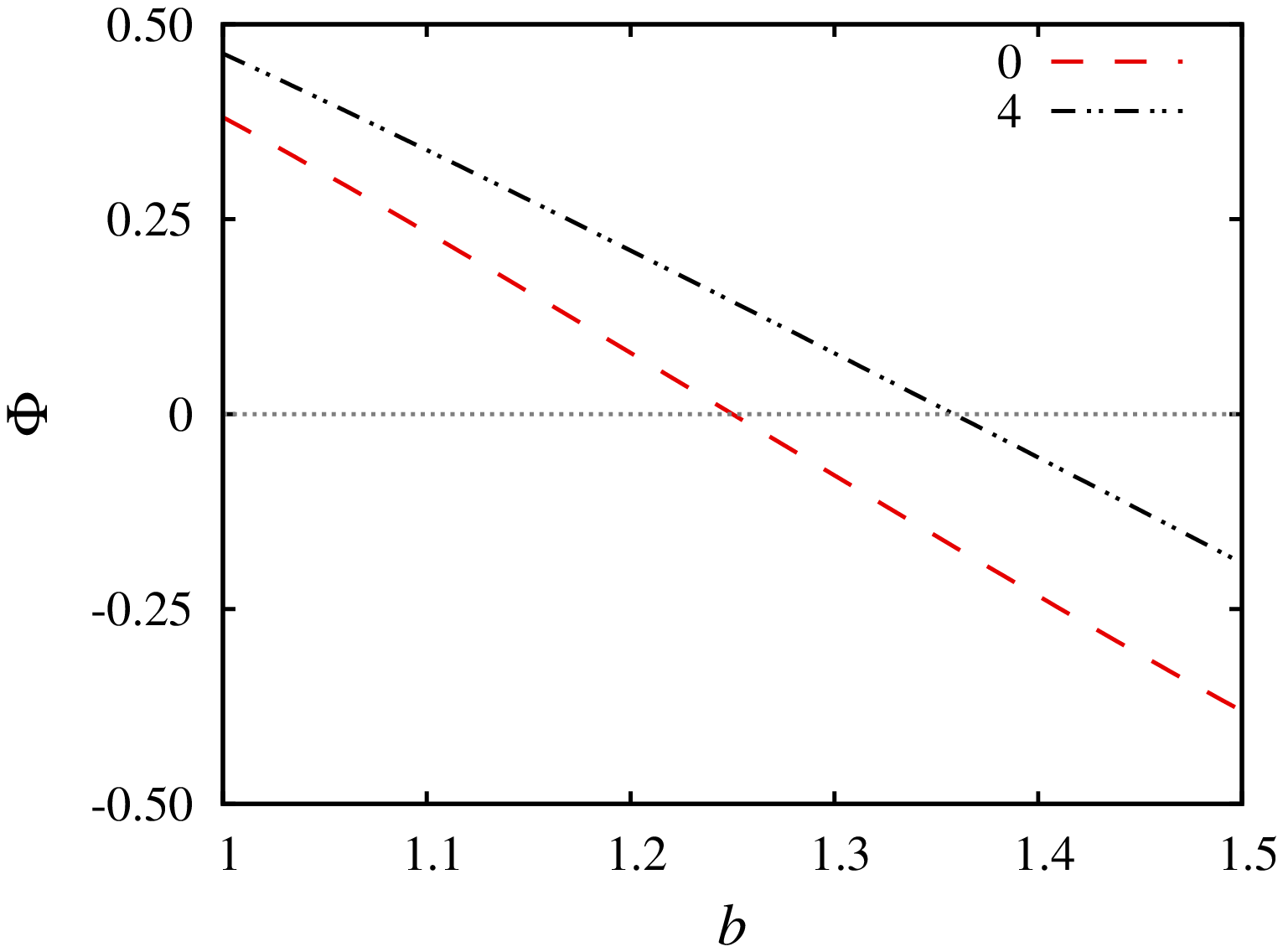,width=8.4cm}}
\caption{\label{leading} Schematic presentation of the leading invasion process and the corresponding difference of invasion probabilities with and without taking into account the ``wisdom of groups''. Arrows in the top two panels depict the leading invasion process for $\alpha=0$ (left) and $\alpha=4$ (right). The invasion that is marked by the gray arrow in the right panel becomes practically irrelevant due to large $\alpha$ (see main text for details). The color code is the same as used in Figs.~\ref{prepared} and \ref{random} to distinguish players with different learning activities. Lower panel depicts the normalized difference of invasion probabilities $\Phi$ of $C \to D$ and $D \to C$ strategy changes in dependence on the temptation to defect $b$ at $K=0.5$. It can be observed that for $\alpha=4$ the invasion front changes sign at a higher value of $b$ than for $\alpha=0$, thus corroborating the reported promotion of cooperation due to the ``wisdom of groups'' at the microscopic level of the evolutionary process.}
\end{figure}

The foundations of this effect can be demonstrated nicely at the microscopic level, where the leading invasion processes that are responsible for the evolution along the interfaces can be studied. Top two panels in Fig.~\ref{leading} show these elementary steps. Due to the introduced neighbor-dependent learning activity, one of the $D \to C$ invasions, marked with the gray arrow, becomes negligible at high $\alpha$ values, in turn leading to a slight bias in the evolutionary process that supports cooperation. This can actually be quantified by calculating the difference of invasion probabilities between $C \to D$ and $D \to C$ transitions for different values of $\alpha$. The bottom panel of Fig.~\ref{leading} features the results, as obtained for $\alpha=0$ and $\alpha=4$. Since the average invasion probability is lowered for higher values of $\alpha$, the difference needs to be normalized by an appropriate factor. More precisely, the plotted quantity $\Phi$ that demonstrates the impact of ``wisdom of groups'' is
\begin{widetext}
\begin{equation}
\Phi = \frac{W(2,2b-2)+W(2,2b-3)-W(2,2-2b)-W(1,3-2b)}
{W(2,2b-2)+W(2,2b-3)+W(2,2-2b)+W(1,3-2b)} \,
\label{crowd}
\end{equation}
\end{widetext}
where
\begin{equation}
W(n,dP) = \left(\frac{n}{z}\right)^{\alpha}  \frac{1}{1+\exp[(dP)/K]} \,
\end{equation}
As the lower panel of Fig.~\ref{leading} shows, the critical $b$ value where the invasion front changes sign is overestimated (see Fig.~\ref{pdsq}) because other processes are neglected by this analysis, yet the fact that this happens at a higher value of $b$ for larger values of $\alpha$ clearly explains the reported promotion of cooperation due to the ``wisdom of groups''. As is well-known, it is beneficial for cooperators to be accumulated while the same is not true for defectors. The revealed dynamics further supports this cause by means of dynamically modified invasion, which is significantly decelerated in the absence of steps along the invasion fronts.

\section*{Discussion}
Summarizing the results, we have shown that by allowing individuals to exploit the ``wisdom of groups'' helps to resolve social dilemmas, regardless of whether they are described by games governed by pairwise interactions or by games governed by group interactions. The main assumption has been that the strategy of each individual player is representative for its knowledge, and that accordingly the aggregate information about this for each particular group is representative for the ``wisdom of the group''. The later has been taken into account by modifying the learning activity of players. If the strategy of a player has been the same as the strategy of the majority of other players within the group, then its propensity to change it, and accordingly its learning activity, would be low. Conversely, if the strategy has been different, the player would be eager to change it, and accordingly it would had a high learning activity. Although the dynamical alteration of learning activities is obviously strategy independent, we have shown that it creates significant evolutionary advantages for the cooperators. Foremost, it slows down the evolutionary process once strategies begin to aggregate. Cooperators are therefore able to build compact clusters with smooth interfaces, while defectors are unable to invade efficiently. On the one hand, the ``wisdom of groups'' thus reinforces the workings of spatial reciprocity \cite{nowak_n92b}, and on the other it prevents defectors to utilize their superior fitness in pairwise comparisons with their neighbors. We have demonstrated this mechanism also at the microscopic level, by comparing the leading invasion process, which is responsible for the evolution along the interfaces, with and without invoking the ``wisdom of groups''. We have concluded that the mechanism of cooperation promotion relies on the spontaneous emergence of a dynamically decelerated invasion process, which is conceptually similar as reported previously in the context of aging in evolutionary games \cite{szolnoki_pre09} and the evolution of public cooperation on interdependent networks \cite{wang_z_epl12}, yet with the crucial difference that here it emerges spontaneously using a completely different motivational and methodological background.

The study of the ``wisdom of crowds'', although formally introduced in the past decade \cite{surowiecki_04}, has roots that go back a whole century \cite{galton_n1907, lorge_pb58}. Although it is accepted that it works in favor of our societies and social welfare \cite{page_07}, it is still critically probed, questioned and elaborated upon even today \cite{rauhut_jmp10, golub_aejm10, lorenz_pnas11}. As we hope this study succeeded to demonstrate, evolutionary games provide a theoretical framework that is very much susceptible to this phenomenon, and we hope to have elicit the interest of readers to continue along this line of research.

\section*{Methods}

We consider two different evolutionary games, namely the prisoner's dilemma and the public goods game, on different lattices of size $L^2$ and degree $z$ with periodic boundary conditions.

The prisoner's dilemma is characteristic for games that are governed by pairwise interactions. It entails cooperators and defectors as the two competing strategies, and it is characterized by the temptation to defect $T = b$, reward for mutual cooperation $R = 1$, and both the punishment for mutual defection $P$ as well as the suckers payoff $S$ equaling $0$. As is standard practice, $1 < b \leq 2$ ensures a proper payoff ranking and captures the essential social dilemma between individual and common interests \cite{nowak_n92b}. Player $x$ with strategy $s_x$ acquires its payoff $\pi_{s_x}$ by playing the game pairwise with all its $z$ neighbors.

The public goods game, on the other hand, is characteristic for games that are governed by group interactions. Accordingly, players are arranged into overlapping groups of size $G$, such that every player is connected to its $z=G-1$ nearest neighbors and is a member of $g=G$ different groups. Here players must decide simultaneously whether they wish to contribute to the common pool or not. All the contributions are then multiplied by a factor $r>1$ to take into account synergetic effects of cooperation, and the resulting amount is divided equally among all $G$ group members irrespective of their initial decision. From the perspective of each individual, defection is clearly the rational decision to make, as it yields the highest personal income if compared to other members of the group. Player $x$ acquires its payoff $\pi_{s_x}^{k}$ by playing the public goods games as a member of the group $k=1 \ldots G$, whereby its overall payoff is thus $\pi_{s_x} = \sum_{k} \pi_{s_x}^{k}$.

Regardless of the interaction network and the type of the game, initially each individual on site $x$ is designated either as a cooperator $s_x=C$ or defector $s_x=D$ with equal probability. The evolutionary process proceeds via the Monte Carlo simulation procedure, comprising the following elementary steps. First, a randomly selected player $x$ acquires its payoff $\pi_{s_x}$. Next, player $x$ chooses one of its nearest neighbors at random, and the chosen player $y$ also acquires its payoff $\pi_{s_y}$ in the same way as previously player $x$. Finally, player $x$ can adopt the strategy $s_y$ from player $y$ with the probability
\begin{equation}
W(s_x \leftarrow s_y) = w_x  \frac{1}{1+\exp[(\pi_{s_x}-\pi_{s_y})/K]} \,
\end{equation}
where $K$ quantifies the uncertainty by strategy adoptions \cite{szabo_pr07} (without loss of generality we use $K=0.1$ for the prisoner's dilemma and $K=0.5$ for the public goods game throughout this paper), and $w_x$ is the learning activity of player $x$. Each full Monte Carlo step (MCS) gives a chance for every player to change its strategy once on average.

\begin{figure}
\centerline{\epsfig{file=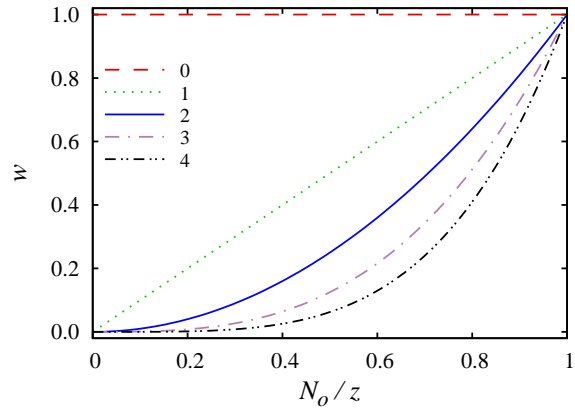,width=8.4cm}}
\caption{\label{alpha} The ``wisdom of groups'' dynamically modifies the learning activity of players. Depicted is the learning activity $w$ in dependence on the fraction of neighboring players that have a different strategy than the focal player, as obtained for different values of $\alpha$ (see figure legend). Traditional versions of both the prisoner's dilemma and the public goods game are recovered if $\alpha=0$, as then the ``wisdom of groups'' is ignored and does not influence the learning activity $w$. On the other hand, for larger values of $\alpha$ the impact of the neighbors becomes increasingly stronger, virtually prohibiting strategy changes that would introduce a strategy different from their own.}
\end{figure}

Importantly, we take the ``wisdom of groups'' into account by defining the learning activity as $w_x=(N_o/z)^\alpha$, where $N_o$ is the number of neighbors that have a different strategy than player $x$ and $z$ is the degree of the lattice. Here the parameter $\alpha$ determines just how seriously the ``wisdom of the group'' is considered. The impact of different $\alpha$ values on the learning activity of players is depicted in Fig.~\ref{alpha}. It can be observed that for $\alpha=0$ the classical version of the evolutionary game is recovered, where the learning activity is independent of the strategies in the neighborhood, and hence the ``wisdom of groups'' is completely ignored. For $\alpha>0$, however, the players are increasingly more prone to stick with the strategy that is representative for their nearest neighbors. In fact, the larger the value of $\alpha$, the stronger the feedback between the group and the individual player.

Presented results were obtained by means of Monte Carlo simulations on lattices of linear size $L=400-1200$. The necessary relaxation times varied between $10^5 - 10^7$. It is worth emphasizing that high values of $\alpha$ significantly decelerate the evolutionary process, which in turn creates the need for employing extremely long relaxation times.

\begin{acknowledgments}
This research was supported by the Hungarian National Research Fund (grant K-101490) and the Slovenian Research Agency (grant J1-4055).
\end{acknowledgments}

\end{document}